\documentclass[aps,prep,epsfile,nofootinbib]{revtex4}
\usepackage{graphics}
\usepackage{slashed}
\usepackage{amssymb}
\usepackage{lscape}
\usepackage{amsmath}
\usepackage{graphicx}
\graphicspath{ {images/} }
\begin{document}
\title{Waves of space-time from a collapsing compact object}
\author{$^{1}$ Jaime Mendoza Hern\'andez\footnote{E-mail: jaime.mendoza@alumno.udg.mx}, $^{3}$ Juan Ignacio Musmarra\footnote{jmusmarra@mdp.edu.ar}, $^{2,3}$ Mauricio Bellini.
\footnote{{\bf Corresponding author}: mbellini@mdp.edu.ar} }
\address{$^1$ Departamento de F\'{\i}sica, Centro Universitario de Ciencias Exactas e Ingenier\'{\i}as, Universidad de Guadalajara
Av. Revoluci\'on 1500, Colonia Ol\'impica C.P. 44430, Guadalajara, Jalisco, M\'exico. \\
$^2$ Departamento de F\'isica, Facultad de Ciencias Exactas y
Naturales, Universidad Nacional de Mar del Plata, Funes 3350, C.P.
7600, Mar del Plata, Argentina.\\
$^3$ Instituto de Investigaciones F\'{\i}sicas de Mar del Plata (IFIMAR), \\
Consejo Nacional de Investigaciones Cient\'ificas y T\'ecnicas
(CONICET), Mar del Plata, Argentina.}

\begin{abstract}
We study the partial time dependent collapse of a spherically symmetric compact object with initial mass $M_1+M_2$ and final mass $M_2$ and the waves of space-time emitted during the collapse via back-reaction effects. We obtain exact analytical solutions for the waves of space-time in an example in which $M_1=M_2=(M_1+M_2)/2$. The wavelengths of the space-time emitted waves during the collapse have the cut (we use natural units $c=\hbar=1$): $\lambda  < (2/b)$, $(1/b)$-being the time scale that describes the decay of the compact object.
\end{abstract}
\maketitle

\section{Introduction and Motivation}

In geometrodynamics\cite{wheeler1,wheeler2} particles and fields are not considered as foreign entities that are immersed in geometry, but are regarded as manifestations of geometry, property. Under certain conditions boundary conditions must be considered in the variation of the action\cite{Y}. In the event that a manifold has a boundary
$\partial{\cal{M}}$, the action should be supplemented by a
boundary term, so that the variational of the action to be
well-defined\cite{GH}.  However, this is not the only manner to
study this problem. As was recently demonstrated\cite{RB}, there
is another way to include the flux that cross a 3D-hypersurface that
encloses a physical source without the inclusion of another term
in the Einstein-Hilbert (EH) action
\begin{equation}
 I_{EH} =  \int d^{4}x \, \sqrt{-g} \left( \frac{R}{2\,\kappa}+\mathcal{L}_{m} \right).
\label{act}
\end{equation}
This is an important fact, because the study of different sources can help to develop a more general and powerful relativistic formalism, where dynamical systems can be studied in a theoretical manner, in order to make a better description of the physical reality. In this work we shall deal only with classical systems, in order to describe waves of space-time. These terms can be seen as a source, described by an extended manifold (with respect to the Riemann background), which has geometrodynamical physical consequences on a gravito-electromagnetic physical system\cite{ruso}.

On the another hand, the understanding of a collapsing system is a very important issue in theoretical physics, mainly in astrophysics and cosmology physics. Some extensive investigations were made related to a spherically symmetric collapse driven by a scalar field \cite{Gundlach}. More recently, this issue has been analytically explored in \cite{GJ,G}. The study of a collapse for a fluid with an heat flux has been treated in \cite{Sharma}. An interesting issue to study is the evolution of the global topology of space time during a collapse driven by a scalar field that can avoid the final singularity. This topic was explored in a recent work \cite{col} jointly with the
geometrical back-reaction of space time produced by this collapse. However, in this work we shall consider a semi-Riemannian manifold, so that $\nabla_{\epsilon} g_{\alpha\beta}=0$.

In this work we are interested in the study of a partial collapse of some compact object with initial mass $M_1+M_2$, that is reduced during this collapse to a final state with mass $M_2$, such that during the transition it transfers a mass $ M_1 \,e^{-b\,t}$ to the space-time, and therefore the collapse generates a spherically symmetric wave of space-time. The paper is organised as follow: in Sect. (\ref{2}) we introduce the physical source for a object that transfers mass to a wave of space-time and we obtain, by varying the action, the new Einstein's equations with sources and the equation for the trace of the space-time wave . In Sect. (\ref{3}) we describe the problem which we are interested in solving, the partial collapse of a spherically symmetric compact object that transfers a part of its mass to the space-time wave. In Sect. (\ref{4}) we propose a particular dynamical evolution for the mass that decreases exponentially with time. We obtain the wave equation for the space-time and we found a exact analytical solution for this equation in a particular case. Finally, in Sect. (\ref{5}) we develop some final comments.

\section{Physical sources from boundary conditions in the variation of the action}\label{2}

We consider the variation of the Einstein-Hilbert (EH) action with respect to the metric tensor
\begin{equation}\label{eqn:AccionGHY}
 \delta I_{EH}  = \frac{1}{2\kappa}\,\int d^{4}x \sqrt{-g} \left[ \delta g^{\alpha \beta} \left( R_{\alpha \beta} - \frac{g_{\alpha \beta}}{2} R + \kappa\, T_{\alpha \beta}\right) + g^{\alpha \beta} \delta R_{\alpha \beta} \right] =0,
\end{equation}
where the stress tensor $T_{\alpha \beta}$ is defined in terms of the variation of the Lagrangian
\begin{equation}
T_{\alpha \beta} = 2\frac{\delta \mathcal{L}_{m}}{\delta g^{\alpha\beta}} - g_{\alpha\beta}\, \mathcal{L}_{m},
\end{equation}
and describes the physical matter fields. By considering a flux $\delta \Phi$ of $\delta W^{\alpha}$ through the 3D closed hypersurface $\partial M$, we obtain that: $g^{\alpha \beta} \delta R_{\alpha \beta} = \nabla_{\alpha}\delta W^{\alpha}=\delta \Phi$, with $\delta W^{\alpha}=\delta
\Gamma^{\alpha}_{\beta\gamma} g^{\beta\gamma}-
\delta\Gamma^{\epsilon}_{\beta\epsilon}
g^{\beta\alpha}=g^{\beta\gamma} \nabla^{\alpha}
\delta\Psi_{\beta\gamma}$\cite{4}. To formalise the study of the system we shall consider boundary conditions that can describe waves of space-time with external sources. In our case the waves of space-time  propagates due to the partial collapse of the spherically compact object, which is the source that produce the waves. In order for describe a dynamical system with a source that generates waves of space-time, we shall consider the case where ${\delta I_{EH}\over \delta S\,\,\,\,\,\,\,\,} =0$ implies that
\begin{equation}\label{vact}
\frac{\delta g^{\alpha\beta}}{\delta S}  \left[ G_{\alpha\beta} + \kappa \,T_{\alpha\beta} \right] + g^{\alpha\beta} \frac{\delta R_{\alpha\beta}}{\delta S\,\,\,\,\,\,}=0 \quad \rightarrow \quad \frac{\delta g^{\alpha\beta}}{\delta S} \underbrace{ \left[ G_{\alpha\beta} + \kappa \,T_{\alpha\beta} \right]}_{=\Lambda \,g_{\alpha\beta}} + \Box \chi = 0,
\end{equation}
that guarantees the preservation of the action, and the recovering of the Einstein's equations with sources. Furthermore, $\chi(x^{\epsilon}) \equiv g^{\mu\nu} \chi_{\mu\nu}$ is a classical scalar field, and $\chi_{\mu\nu}={\delta \Psi_{\mu\nu}\over \delta S\,\,\,\,\,\,\,\,}$ describes the waves of space-time produced by the source through the 3D hypersurface. Therefore, the condition ${\delta I_{EH}\over \delta S\,\,\,\,\,\,\,\,} =0$, with boundary conditions included, in the case described by (\ref{vact}), can be written as
\begin{eqnarray}
{G}_{\alpha \beta} & + & \kappa \,T_{\alpha \beta}= \Lambda \, \bar{g}_{\alpha\beta}, \label{aa} \\
\Box \chi &=&  \frac{\delta \Phi}{\delta S}= \Lambda \, g^{\alpha\beta} \frac{\delta g_{\alpha\beta}}{\delta S\,\,\,\,\,\,}, \label{bb}
\end{eqnarray}
such that $g^{\alpha\beta} {\delta g_{\alpha\beta}\over  \delta S\,\,\,\,}  = [1/(-g)] \frac{d(-g)}{dS\,\,}= \frac{d}{dS} \left[{\ln}(-g)\right]$, and $(-g)$ is the positive value of the determinant for the metric that describes the source: $g_{\mu\nu}$.
Therefore, the variation of the flux that cross the 3D-gaussian hypersurface with respect to the line element $S$, is described by
\begin{equation}
\frac{\delta \Phi}{\delta S} = \Lambda \, g^{\alpha\beta} \frac{\delta g_{\alpha\beta}}{\delta S\,\,\,\,\,\,} = \Lambda \, \frac{d}{dS} \left[{\ln}(-g)\right],
\end{equation}
that is the source of the wave equation (\ref{bb}). Here, the cosmological constant $\Lambda$ is related with the flux $\Phi$ and the volume of the manifold: $\sqrt{-g}$:
\begin{equation}
\Lambda =\frac{1}{2} \frac{d\Phi\,\,\,\,\,\,\,\,\,\,\,\,\,\,\,\,\,\,\,\,}{d\left[\ln\left[\sqrt{-g}\right]\right]},
\end{equation}
where we have made use of the fact that $g^{\alpha\beta} {\delta g_{\alpha\beta}\over  \delta S\,\,\,\,} = \frac{d}{dS} \left[{\ln}(-g)\right]=
2\frac{d}{dS} \left[{\ln}(\sqrt{-g})\right]$. When $\frac{\delta \Phi}{\delta S} $ is linear with $\frac{d}{dS} \left[{\ln}(-g)\right]$, hence $\Lambda$ is a constant. This is the case we shall consider in this work.

\section{Partial collapse of a compact object}\label{3}

In order to consider the dynamical collapse of a spherically symmetric compact object where the mass evolves in agreement with (\ref{masa}), we must consider the initial Schwarzschild metric of the object with mass $M_1+M_2$
\begin{equation}\label{m1}
dS^2 = \bar{g}_{\alpha\beta}\,dx^{\alpha}\,dx^{\beta} = f(r) \,dt^2 - \frac{1}{f(r)}\, dr^2 - r^2\, d\Omega^2,
\end{equation}
where $f(r)= 1- 2G(M_1+M_2)/r$ and $d\Omega^2 = d\theta^2+\sin^2(\theta) \,d\phi^2$. During the collapse the mass of the compact object is dynamically reduced from $(M_1+M_2)$ to $M_2$, so that the difference of mass is transferred to the wave of space-time. The transition between both states is described by the function $M(t)$, such that $M(t=0)=M_1+M_2$ and $\lim_{t\rightarrow \infty}M(t)\rightarrow M_2$. Furthermore, $f(r) \rightarrow f(r,t)$ and the metric during the transition is:
\begin{equation}\label{m2}
dS^2 = {g}_{\alpha\beta}\,dx^{\alpha}\,dx^{\beta} =  f(r,t)\, dt^2 - a^2(t)\left[\frac{dr^2}{f(r,t)}  - r^2\, d\Omega^2\right],
\end{equation}
with $f(r,t)=1- 2G M(t)/r$ and $a(t)\leq a_0$ is the radius of the compact object, which will be considered as proportional to the time dependent mass: $a(t)=C\, G\,M(t)$. Here, $C={a_0\over G (M_1+M_2)} >2$ is a constant, such that $a_0$ is the initial radius of the compact object. Notice that (\ref{m2}) is the metric that describes the physical source of the wave of space-time and (\ref{m1}) is the metric that characterizes the initial state on which will propagates the wave produced by the source. At the end of the collapse the remaining mass is $M_2=\lim_{t\rightarrow \infty} M(t)$, and the residual metric will be
the same than (\ref{m1}), but with a mass $M_2$. It is expected that, during the collapse, where the compact object losses a mass $\Delta M=M_1$. In absence of dissipative effects, it is expected that the lost energy by the star $\Delta E=M_1\,c^2$, be transferred to the wave of space-time, that will spread at the speed of light as a spherical wave.

The equation of state : $\omega(t,r)=p(t,r)/\rho(t,r)$ (here $p$ and $\rho$ are respectively pressure and energy density), for this spherically symmetric compact object that has a radius $a(t)=C\,M(t)$, that decreases with time, is
\begin{equation}\label{om}
\omega(t,r)=\frac{\Lambda \,M^2 \left[4\left((MG)^2-GM\,r\right) + r^2\right] -2 r \ddot{M} M \left[r-2\,GM\right] - \dot{M}^2 r^2}{\Lambda\,M^2 \left[r^2-4\,GM \left(r-GM\right) \right] - \dot{M}^2\left(4GMr - 3r^2\right)},
\end{equation}
where ${a_0\over (M_1+M_2)}\,M(t) > r\geq 2\,G\, M(t)$, with ${a_0\over G(M_1+M_2) }>2$. It is expected that during the collapse $\omega>0$.

\section{An example}\label{4}

To examine the theory, we shall consider the case where the evolution of the mass is
\begin{equation}\label{masa}
M(t) = M_1 \,e^{-b\,t} + M_2,
\end{equation}
where $t\geq 0$ is the time, $b$ is some positive constant such that $1/b$ give us the scale of decaying of the compact object. To evaluate the equation of state inside the star, we shall consider a radius $r= \alpha \,G M(t)$ which is smaller than the stellar radius and bigger than the dynamic Schwarzschild one: $ 2 <\alpha\leq {a_0\over G (M_1+M_2)} $
For the case (\ref{masa}), the equation of state of the system for $a(t)= {a_0\over (M_1+M_2)}\,M(t)$, is
\begin{equation}
\omega(t,r)=\frac{\Lambda \left[M^2 (\alpha - 2)^2\right] - \left[2 M \ddot{M} (\alpha - 2) + \alpha \dot{M}^2\right]}{\Lambda \left[M^2 (\alpha - 2)^2\right]+\alpha (3\alpha - 4) \dot{M}^2}.
\end{equation}
Notice that the equation of state in the limit case where $t\rightarrow \infty$ is
\begin{equation}
\lim_{t\rightarrow \infty} \omega(t,r) \rightarrow 1.
\end{equation}
Furthermore, in the limit case where $\alpha \rightarrow 2$, we obtain the equation of state
\begin{equation}
\lim_{\alpha\rightarrow 2} \omega(t,r) \rightarrow -\frac{1}{2}.
\end{equation}

\begin{figure}
  \centering
    \includegraphics[width=13cm, height=14cm]{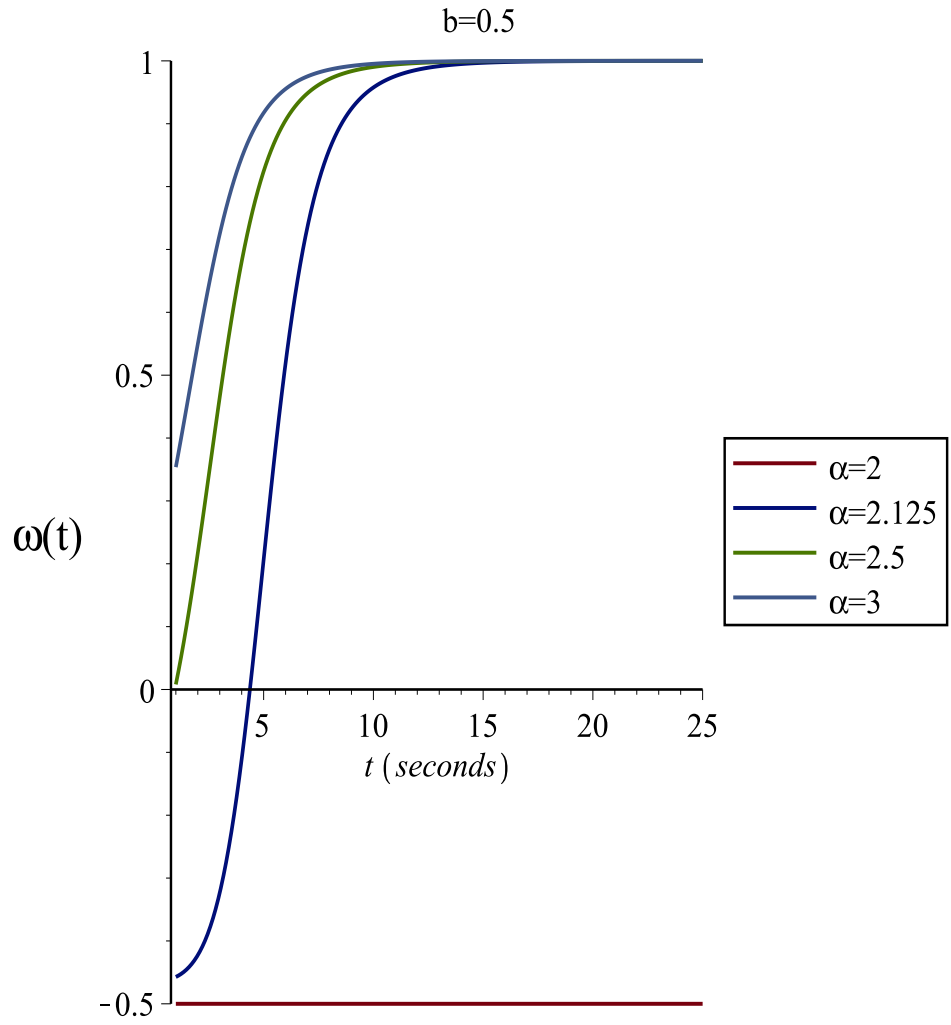}
  \caption{Plotting of the $\omega(r,t)$ as a function of $t$ (seconds), for different $\alpha$-values corresponding to different $r$-values, with $M_1=M_2=0.5\,M_{\bigodot}$.  For simplicity, we have done $M_{\bigodot}=1$. Notice that on the Schwarzschild radius $\omega(\alpha=2)=-1/2$, but for different (and larger) $\alpha$-values, the equation of state tends asymptotically to $\omega\rightarrow 1$ as the collapse evolves.}
  \label{F1}
\end{figure}

\begin{figure}
  \centering
    \includegraphics[width=13cm, height=14cm]{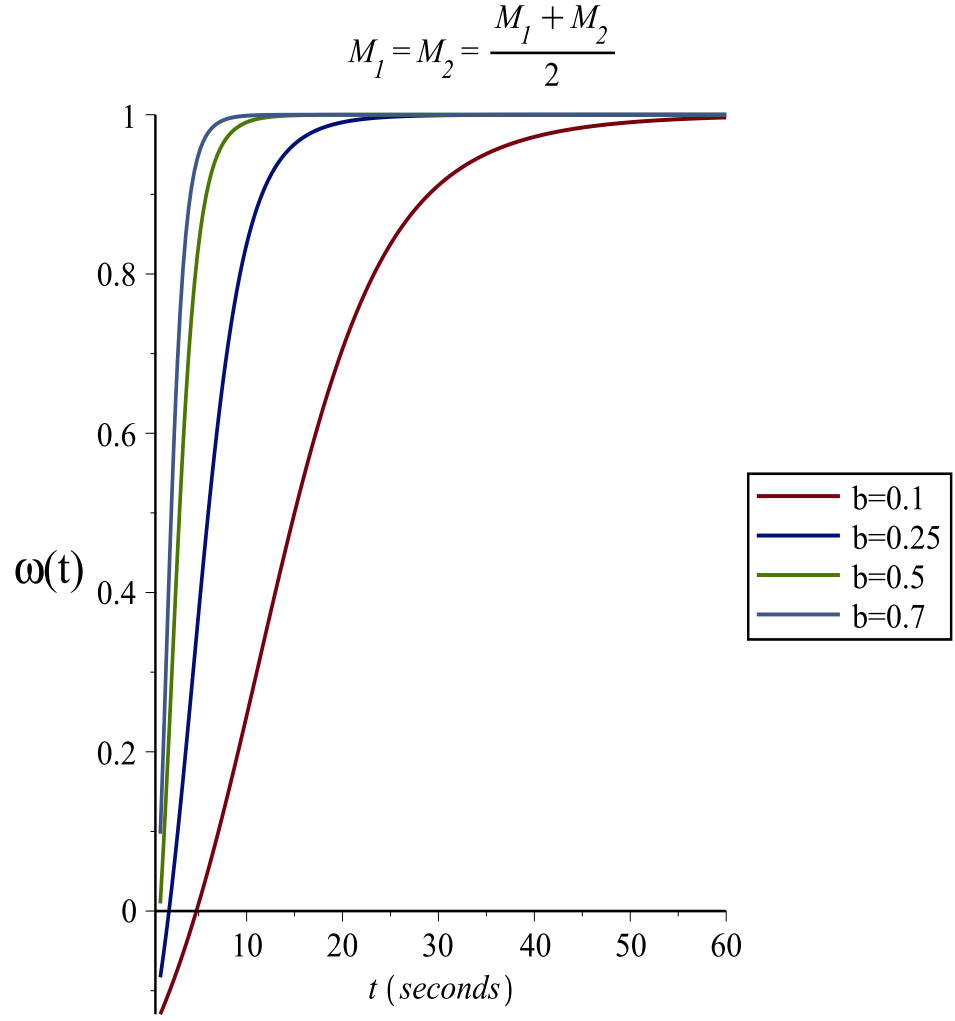}
  \caption{Plotting of the $\omega(r,t)$ as a function of $t$ (seconds), for different $b$-values and $M_1=M_2=0.5\,M_{\bigodot}$ and $\alpha=2.5$. For simplicity, we have done $M_{\bigodot}=1$. Notice that initially $\omega < 0$, later $\omega\rightarrow 1$ as the collapse evolves.}
  \label{F2}
\end{figure}

In the figure (\ref{F1}) we have plotted $\omega(r,t)$ as a function of $t$, for different $\alpha$-values corresponding to different $r$-values of the interior of the star. We have considered $M_1=M_2=0.5\,M_{\bigodot}$. Notice that on the Schwarzschild radius $\omega(\alpha=2)=-1/2$, but for different (and larger) $\alpha$-values, the equation of state tends asymptotically to $\omega\rightarrow 1$ as the collapse evolves. In the figure (\ref{F2})
we have plotted $\omega(r,t)$ as a function of $t$ for different $b$-values, using $M_1=M_2=0.5\,M_{\bigodot} $ and $\alpha=2.5$. Notice that initially $\omega < 0$, but later $\omega\rightarrow 1$ as the collapse evolves.

We can write the complete solution for the space-time wave described by the field $\chi(t,r,\theta,\phi)$, as
\begin{equation}
\chi(t,r,\theta,\phi)= \sum_{n=0}^{\infty}\,\chi_n(t,r,\theta,\phi).
\end{equation}
For modes that include waves of gravitational origin, must be with $l=2$, and $-2\leq m\leq 2$, so that
\begin{eqnarray}
\chi_n(t,r,\theta,\phi)= \,\sum_{m =-2}^{2} \left[A_{n,l,m}\,\chi_{n,l,m}(t,r,\theta,\phi)+ A^{\dagger}_{n,l,m}\,\chi^*_{n,l,m}(t,r,\theta,\phi)\right],
\end{eqnarray}
such that the modes $\chi_{n,l,m}(t,r,\theta,\phi)$, are:
\begin{equation}
\chi_{n,l,m}(t,r,\theta,\phi) = \left(\frac{E_{n,l}}{\hbar}\right)^2 \, R_{n,l}(r)\, Y_{l,m}(\theta,\phi)\, {\tau}_n(t),
\end{equation}

We can expand the field $\chi$ as a superposition $\chi_{n,l,m}(t,r,\theta,\phi) \sim R_{n,l}(t,r)\,Y_{l,m}(\theta,\phi)$, where the functions $Y_{l,m}(\theta,\phi)$ are the usual spherical harmonics
\begin{equation}
Y_{l,m}(\theta,\phi)= \sqrt{\frac{(2l+1)(l-m)!}{4\pi\,(l+m)!}} \, {\cal P}^l_m(\theta)\,e^{i\,m\phi},
\end{equation}
where $!$ denotes the factorial and ${\cal P}^l_m(\theta)$ are the Legendre polinomials. In this work we shall consider $l=2$ because we are dealing with waves of space-time and $-2\leq m\leq 2$.

\subsection{Waves of space-time}

We must calculate the equation (\ref{bb}), which in our example takes the form
\begin{equation}\label{eeq}
g^{\alpha\beta} \nabla_{\alpha} \nabla_{\beta} \chi(t,r,\theta,\phi) = \frac{6\,\dot{M}(t)\,r}{M(t)\left(r-2G\,M(t)\right)} \Lambda,
\end{equation}
where we have made use of the fact that ${d\over dS}=U^{\alpha} {d\over d\,x^{\alpha}}$. Notice that all the energy transferred by the compact object to the wave is $M_1\,c^2$, so that the total energy of the system (star $+$ space-time wave), is conserved. For a co-moving observer $U^i=0$ and $U^0=\sqrt{g^{00}}$ in the metric (\ref{m2}), because we are describing the source of the wave in the right side of the equation (\ref{eeq}). In order to calculate the covariant derivatives in the D'Alambertian of the left side in (\ref{eeq}), we must use the connections of the
metric (\ref{m1}), with a mass $M_1+M_2$. This is because when the collapse begins, the mass of the compact object is $M_1+M_2$ and the information that changes the space-time travels with the wave of space-time. Hence, in a given point of the space-time, an observer will feel a gravitational field due to the mass $M_1+M_2$ before the wave leads to the observer, and a gravitational field due to the residual mass $M_2$, after the wave has passed the observer. The equation (\ref{eeq}), written explicitly, is
\begin{eqnarray}
&-&\frac{2}{r}\,{\frac {\partial \chi_{n,l,m}}{\partial r \,\,\,\,\,\,}} -{\frac {
\partial ^{2}\chi_{n,l,m}}{\partial {r}^{2}\,\,\,\,\,\,\,\,\,}} +{\frac{2G\bar{M}}{{r}^{2}}   {\frac {\partial \chi_{n,l,m}}{\partial r \,\,\,\,\,\,}}
   } +
{\frac{2 G\bar{M}}{r}  {\frac {\partial ^{2}\chi_{n,l,m}}{\partial {r}^{2}\,\,\,\,\,\,}} }
\nonumber \\
&-& \frac{1}{{r}^{2}}{\frac {\cos\left( \theta \right)}{{\sin \left( \theta \right)}}  {\frac {\partial \chi_{n,l,m}}
{\partial \theta \,\,\,\,\,\,}}   }
 -{\frac{1}{{r}^{2}  \sin^2 \left(
\theta\right)   } {\frac {\partial ^{2}\chi_{n,l,m}}{\partial {\phi}^{2
}\,\,\,\,\,\,}} }
-\frac {r}{\left[2\,G\bar{M}-r\right]}        { \frac {\partial ^{2}\chi_{n,l,m}}{
\partial {t}^{2}\,\,\,\,\,\,}} -{\frac{1}{{r}^{2} \sin^2 \left( \theta \right) } {\frac {\partial ^{2}\chi_{n,l,m}}{\partial {\theta}^{2}\,\,\,\,\,\,}} } \nonumber \\
&+& {
\left[\frac { 6\, {\dot{M}(t)}\,
  r}{M(t)   \left[ 2\,GM(t)
-r  \right] }\right] \Lambda}=0,
\end{eqnarray}
where we have denoted $\bar{M}=(M_1+M_2)$. We can make variables separation: $\chi_{n,l,m}(t,r,\theta, \phi)=R_{n,l}(t,r)\,Y_{l,m}(\theta,\phi)$, and after making $r=\alpha\,M(t)$ in the source with the approximation of it, for sufficiently large time, we obtain the differential equations:
\begin{eqnarray}
&\,& {2 r}\,{\frac {\partial R_{n,l}(t,r)}{\partial r \,\,\,\,\,\,}} + r^2 {\frac {
\partial ^{2}R_{n,l}(t,r)}{\partial {r}^{2}\,\,\,\,\,\,\,\,\,}} -{{2G\bar{M}}   {\frac {\partial R_{n,l}(t,r)}{\partial r \,\,\,\,\,\,}}
   } -
{{2 G\bar{M}\,r}  {\frac {\partial ^{2}R_{n,l}(t,r)}{\partial {r}^{2}\,\,\,\,\,\,}} }  \nonumber \\
&+&\frac{r^3}{[2G\bar{M}-r]}\frac{\partial^2 R_{n,l}(t,r)}{\partial t^2}- l\,(l+1)\,R_{n,l}(t,r) =  \frac{3 \bar{M}^2  e^{-b\,t} \alpha^3 \,\Lambda}{(\alpha-2)}, \label{uno} \\
&\,& {\frac {\cos\left( \theta \right)}{{\sin \left( \theta \right)}}  {\frac {\partial Y_{l,m}(\theta,\phi)}
{\partial \theta \,\,\,\,\,\,}}   }
 + \frac{1}{ \sin^2 \left(
\theta\right)}    {\frac {\partial ^{2} Y_{l,m}(\theta,\phi)}{\partial {\phi}^{2
}\,\,\,\,\,\,}}
 +  \frac{1}{ \sin^2 \left( \theta\right)} \left( \theta \right)  {\frac {\partial ^{2} Y_{l,m}(\theta,\phi)}{\partial {\theta}^{2}\,\,\,\,\,\,}} \nonumber \\
&+& l\,(l+1)\,Y_{l,m}(\theta,\phi)=0.
\end{eqnarray}
The solution for Eq. (\ref{uno})
\begin{eqnarray}
R_{n,l}(t,r)&=& \left(\frac{6}{\alpha-2} \right) {{\rm e}^{b \left( r-t \right) }}{(G\bar{M})}^{3}b{\alpha}^{3}\Lambda\,
 \left( 2\,G\bar{M}-r \right) ^{2\,bG\bar{M}} {\it H_C}^{(1)} \nonumber \\
 &\times & \int \!\frac{{{\rm e}^{-br}}{\it H_C}^{(2)}
 \left( 2\,G\bar{M}-r \right)^{-2\,bG\bar{M}}}{{r} \left( 8\,{\it H_C}^{(1)}  {\it H_C}^{(2)}  b{(G\bar{M})}^{2}-2\,{\it H_C}^{(1)}  {\it H_{CP}}^{(2)}   G\bar{M}+ {\it H_C}^{(1)}  {\it H_{CP}}^{(2)}
 r+2\,{\it H_C}^{(2)}  {\it H_{CP}}^{(1)}  G\bar{M}-{
\it H_C}^{(2)}  {\it H_{CP}}^{(1)}r\right)}{dr}   \nonumber \\
&-& \left(\frac{6}{\alpha-2} \right)\,{{\rm e}^{b \left( r-t \right) }}{(G\bar{M})}^{3}b{
\alpha}^{3}\Lambda\, \left( 2\,G\bar{M}-r \right) ^{-2\,bG\bar{M}}  {\it H_C}^{(2)} \nonumber \\
&\times & \int \! \frac{\left( 2\,
G\bar{M}-r \right) ^{2\,bG\bar{M}}{\it H_C}^{(1)}  {{\rm e}^{-br}}}{ {r} \left( 8\,{\it H_C}^{(1)}  {\it H_C}^{(2)}  b{(G\bar{M})}^{2}-2\,{
\it H_C}^{(1)}  {\it H_{CP}}^{(2)}  G\bar{M}+{\it H_C}^{(1)}  {\it H_{CP}}^{(2)}  r+2\,{\it
H_C}^{(2)}  {\it H_CP}^{(1)}  G\bar{M}-{\it H_C}^{(2)}  {\it H_{CP}}^{(1)}  r \right)}{dr}, \nonumber \\
\end{eqnarray}
where we have denoted the confluent Heun functions ${\it H_C}$ and the prime confluent Heun functions ${\it H_{CP}}$, as
\begin{eqnarray}
{\it H_C}^{(1,2)} &=& {\it H_C} \left( -4\,bG\bar{M},\mp 4\,bG\bar{M},0,8\,{b}^{2}{(G\bar{M})}^{2},-8\,{b}^{2}{(G\bar{M})}^{2}-l(l+1)
,\,{\frac {2\,G\bar{M}-r}{2 G\bar{M}}} \right), \\
{\it H_{CP}}^{(1,2)} & =& {\it H_CP} \left( -4\,bG\bar{M},\mp 4\,bG\bar{M},0,8\,{b}^{2}{(G\bar{M})}^{2},-8
\,{b}^{2}{(G\bar{M})}^{2}-l(l+1),\,{\frac {2\,G\bar{M}-r}{2G\bar{M}}} \right).
\end{eqnarray}
Notice that we have made null the constants in the homogenous contributions to the solution for $R_{n,l}(t,r)$ because it is expected that in absence of sources, there will no emission of space-time waves.

\begin{figure}
  \centering
    \includegraphics[width=13cm, height=14cm]{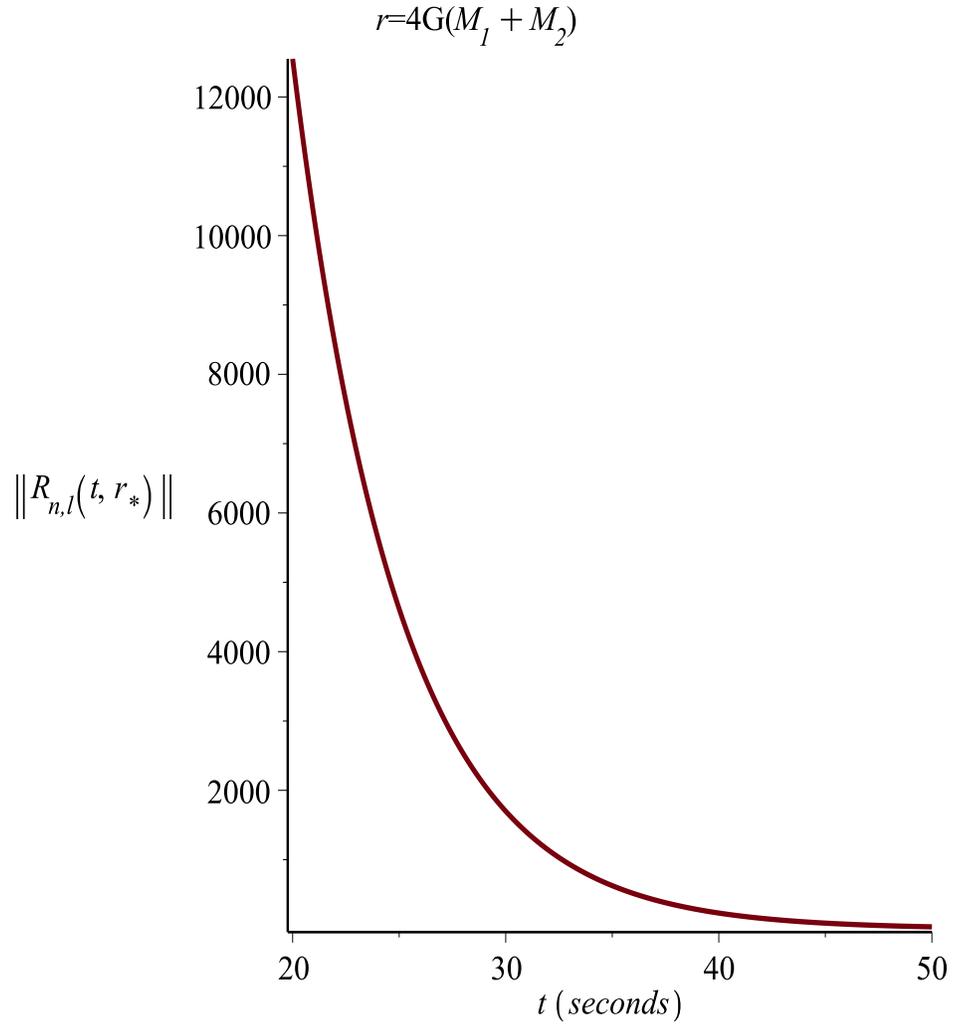}
  \caption{Plotting of norm $|R_{n,l}(t,r_*)|$ as a function of $t$ (seconds), with $M_1=M_2=2.5\,M_{\bigodot}$, for an observer (which is outside the star), at a distance $r_*=4\,G(M_1+M_2)$. We have considered $\Lambda=3/(b^2)$, with $b=0.2$ and $a_0=2.2\,G(M_1+M_2)$. For simplicity, we have done $M_{\bigodot}=1$.}
  \label{F3}
\end{figure}

\begin{figure}
  \centering
    \includegraphics[width=13cm, height=14cm]{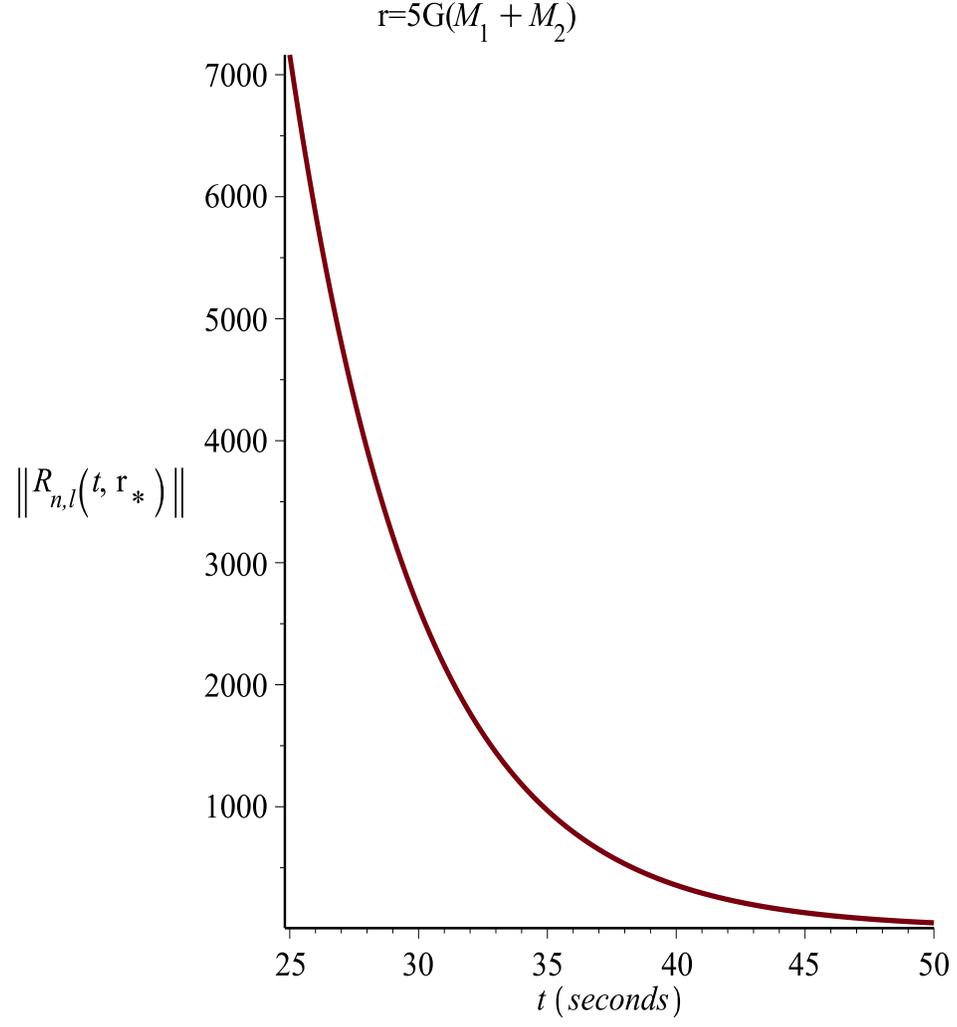}
  \caption{Plotting of norm $|R_{n,l}(t,r_*)|$ as a function of $t$ (seconds), for $M_1=M_2=2.5\,M_{\bigodot}$ and $r_*=5\,G(M_1+M_2)$. We have considered $\Lambda=3/(b^2)$, with $b=0.2$ and $a_0=2.2\,G(M_1+M_2)$. For simplicity, we have done $M_{\bigodot}=1$.}
  \label{F4}
\end{figure}

In the figures (\ref{F3}), and (\ref{F4}) we have plotted the norm of $R_{n,l}(t,r_*)$ as a function of time, for different values of $r_*$, and $M_1=M_2=2.5\,M_{\bigodot}$.
Here, $r_*$ is the distance where is situated an observer outside the star. We have considered $\Lambda=3/(b^2)$, with $b=0.2$, $a_0=2.2\,G(M_1+M_2)$ and $l=2$, because we are dealing with waves of space-time. In the figure (\ref{F3}) the observer is close to the star (at $r_*=4\,G(M_1+M_2)$), but in (\ref{F4}) is more distant (at $r_*=5\,G(M_1+M_2)$). Notice that the signal is more weak in this last. Because the wavelengths must be smaller than the initial size of the star, the wavelengths of the emitted space-time waves, must be
\begin{equation}
\lambda \equiv \frac{2\pi}{k} < a_0 =(2/b),
\end{equation}
where $(1/b)$ is the time scale that describes the decay rate of the compact object. Finally, notice that the equation of state change its signature during the collapse in the interior of the compact object for different interior radius, but no for a Schwarzschild one.

\section{Final comments}\label{5}

We have studied the dynamics of a partial collapse for a spherically symmetric compact object and the waves of space-time emitted during the collapse when the collapsing body
delivers a part of its mass during the collapse. Here, the waves of space-time are described by back-reaction effects. To study these effects we have modified the boundary conditions in the minimum action principle by considering a dynamical source through a 3D gaussian wave-like flux that cross such closed hypersurface. The effective dynamics is described by the system of equation (\ref{aa}) and (\ref{bb}), where $\chi$ is the trace of the wave components: $\chi_{\alpha\beta}={\delta \Psi_{\alpha\beta}\over \delta S\,\,\,}$ on the background curved space-time. Notice that we are dealing with waves that do not comes from the variation of a quadrupole momentum, but distortions of space-time which propagates as waves due to the collapse of a spherically symmetric compact object. The description of the dynamics is extremely complicated, but, fortunately we have founded exact analytical solutions in an example for which the final mass is $M_2=(M_1+M_2)/2$. The wavelengths of the space-time emitted waves during the collapse have the cut: $\lambda  < (2/b)=a_0$, where $(1/b)$ is the time scale that describes the decay rate of the compact object. Notice that the approach here used is different to other used in some previous works\cite{col,jmc1} in which boundary conditions are related to a displacement from a semi-Riemannian manifold to another extended manifold in which $\nabla_{\epsilon} g_{\alpha\beta}\neq 0$, to describe back-reaction effects. However, in this work we have used a semi-Riemannian manifold to describe boundary conditions, so that $\nabla_{\epsilon} g_{\alpha\beta}=0$. This is due to the fact, in this work we have considered that the energy lost through the compact object is transferred to the wave of space-time.

\section*{Acknowledgements}
\noindent
J. M. H. acknowledges CONACYT (M\'exico) for fellowship support ("Beca de movilidad al extranjero"). J. I. M. and M. B. acknowledge CONICET, Argentina (PIP 11220150100072CO) and UNMdP (EXA852/18) for financial support. This research was supported by the CONACyT-UDG Network Project No. 294625 "Agujeros Negros y Ondas Gravitatorias".

\newpage

\end{document}